\PassOptionsToPackage{table,xcdraw}{xcolor}
\documentclass[conference]{IEEEtran}
\usepackage[table,xcdraw]{xcolor}
\usepackage{tikz}
\usepackage{tikzpagenodes}
\usepackage{algorithm}
\usepackage{algpseudocode}
\usepackage{tcolorbox}
\usepackage{fontawesome}
\tikzset{
  startstop/.style={rectangle, rounded corners, minimum width=3cm, minimum height=1cm, text centered, draw=black, fill=white!30},
  process/.style={rectangle, minimum width=3cm, minimum height=1cm, text centered, draw=black, fill=white!30},
  subprocess/.style={rectangle, minimum width=3cm, minimum height=1cm, text centered, draw=black, fill=white!30},
  arrow/.style={thick,->,>=stealth},
}
\def\BibTeX{{\rm B\kern-.05em{\sc i\kern-.025em b}\kern-.08em
    T\kern-.1667em\lower.7ex\hbox{E}\kern-.125emX}}

\usepackage{amssymb}
\usepackage[switch]{lineno}
\usepackage{hyperref}
\usepackage{color}
\usepackage{multirow}

\usepackage[normalem]{ulem}
\usepackage{xcolor}

\usepackage{booktabs} 
\usepackage{amssymb}
\usepackage{mathtools}
\usepackage{amsmath}
\usepackage{blkarray}
\usepackage{rotating} 
\usepackage[table,xcdraw]{xcolor}
\usepackage{makecell}
\usepackage{pifont}%
\usepackage{footmisc}
\usepackage{soul}
\usepackage{bbding}
\usepackage{enumerate}
\usepackage{soul} % for highlighting text

\usepackage{subcaption}

\usepackage{multirow}
\usepackage{enumitem}

\begin{document}

%\title{Integrating AoI-Aware UAV and IRS for Enhanced Mobile Network Security}
\title{Securing the Skies: An IRS-Assisted AoI-Aware Secure Multi-UAV System with Efficient Task Offloading}
%\title{Task Optimization and Security Enhancement in UAV Networks: A Transformer-DRL Approach with IRS Assistance}

% \author{Joshi Poorvi,~\IEEEmembership{Student Member,~IEEE,}  Alakesh Kalita,~\IEEEmembership{Member,~IEEE,}  Mohan Gurusamy,~\IEEEmembership{Senior Member,~IEEE}

\author{\IEEEauthorblockN{$^1$ Joshi Poorvi,  $^2$Alakesh Kalita, $^3$Mohan Gurusamy}
\IEEEauthorblockA{$^{1,3}$Electrical and Computer Engineering, National University of Singapore, Singapore\\
$^{2}$ISTD Pillar, Singapore University of Technology and Design, Singapore\\
 $^1$e1144005@u.nus.edu, $^2$alakesh\_kalita@sutd.edu.sg, $^3$gmohan@nus.edu.sg}

    % \thanks{J.~Poorvi and  M.~Gurusamy are with the Communications \& Networks Lab, Department of Electrical and Computer Engineering, National University of Singapore. (e-mail: \texttt{e1144005@u.nus.edu}, \texttt{gmohan@nus.edu.sg})}

    % \thanks{A.~Kalita is with the ISTD Pillar, Singapore University of Technology and Design, Singapore (e-mail: \texttt{alakesh\_kalita@sutd.edu.sg})}

}
% \markboth{IEEE Internet of Things, ~Vol.~X, No.~X, June~2023}%
% {Hazra \MakeLowercase{\textit{et al.}}: Bare Demo of IEEEtran.cls for IEEE Journals}
\maketitle

\thispagestyle{empty} % Remove page number from first page

\pagestyle{empty} % Remove page numbers from all pages

\begin{abstract}

Unmanned Aerial Vehicles (UAVs) are integral in various sectors like agriculture, surveillance, and logistics, driven by advancements in 5G. However, existing research lacks a comprehensive approach addressing both data freshness and security concerns. In this paper, we address the intricate challenges of data freshness, and security, especially in the context of eavesdropping and jamming in modern UAV networks. Our framework incorporates exponential AoI metrics and emphasizes secrecy rate to tackle eavesdropping and jamming threats. We introduce a transformer-enhanced Deep Reinforcement Learning (DRL) approach to optimize task offloading processes. Comparative analysis with existing algorithms showcases the superiority of our scheme, indicating its promising advancements in UAV network management.
\end{abstract}

\begin{IEEEkeywords}
Unmanned Aerial Vehicles, Age of Information, Intelligent Reflecting Surfaces, Deep Reinforcement Learning, Physical Layer Security, Data Freshness, Task Offloading.
\end{IEEEkeywords}

%\tableofcontents

\section{Introduction}
\label{sec:introduction}

Unmanned Aerial Vehicles (UAVs), have seen increased utilization in agriculture, surveillance, logistics, and emergency response, facilitated by advancements in 5G networks, providing high-speed, low-latency communication.
In UAV networks, maintaining data freshness is vital, especially in emergencies such as natural disasters or search and rescue operations. Data freshness refers to the timeliness of information collected by UAVs, closely linked to the Age of Information (AoI), measuring the duration between data acquisition and availability for decision-making. Recent studies underscore the importance of AoI in UAV networks. \cite{choudhury2021aoi} emphasizes the importance of minimizing AoI for timely information collection and processing in UAV-aided Mobile Edge Computing Networks. Similarly, \cite{rahimi2023minimizing} investigates trajectory planning for multiple UAVs in IoT networks to minimize average AoI. Additionally, \cite{9426899} utilizes AoI as a metric to assess temporal correlation in IoT data packets and proposes an AoI-energy-aware data collection scheme for UAV-assisted IoT networks. These studies underscore the significance of reducing overall AoI to maintain data freshness below a specified threshold.
Furthermore, Security threats such as eavesdropping and jamming pose significant challenges to UAV network reliability and integrity. Various Physical Layer Security (PLS) methods, including multi-antenna relaying and artificial noise, aim to mitigate these threats \cite{wei2023secure}. These strategies aim to mitigate eavesdropping channels and reduce the interception of information by unauthorized receivers. Recent research has explored the use of Intelligent Reflecting Surfaces (IRS) to enhance PLS in UAV networks. IRS dynamically adjusts reflecting elements to improve signal reception by legitimate users while reducing unauthorized channel quality \cite{shakhatreh2023mobile}. While previous studies have focused on individual optimization of UAV-IRS integration for AoI metrics \cite{10092806}, but simultaneous optimization of secrecy rate and AoI metrics in a unified framework remains unexplored. There's a need for a comprehensive framework addressing security challenges related to jamming, eavesdropping, and data freshness. 

Our work focuses on optimizing an IRS-assisted AoI aware bi-layer multi-UAV system. A cooperative data sensing and transmission framework is developed where UAVs utilize radar signals to gather status information from users and securely relay it to the base station in the presence of eavesdroppers and jammers. The optimization challenge involves jointly considering the trajectory of the UAVs and the IRS reflection vector for PLS. The main contribution includes devising strategies to optimize the task offloading process within the proposed system, including designing non-overlapping trajectories for Computational-UAVs (C-UAVs) at the lower layer and IRS-aided UAVs (I-UAVs) operating at higher altitudes, focusing on collision avoidance and determining optimal beamforming vectors for I-UAVs. Further, a framework is introduced that incorporates exponential penalty-based AoI metrics and overall secrecy rate, aiming to enhance data freshness and security within network. At last, we used Multi-agent Deep Reinforcement Learning (DRL) supported by Gated Transformer architecture for joint optimization, facilitating efficient temporal modeling to optimize across multiple agents.

\section{System and Channel Model}
\begin{figure}[!t]
    \centering 
    \includegraphics[width=1.1\linewidth]{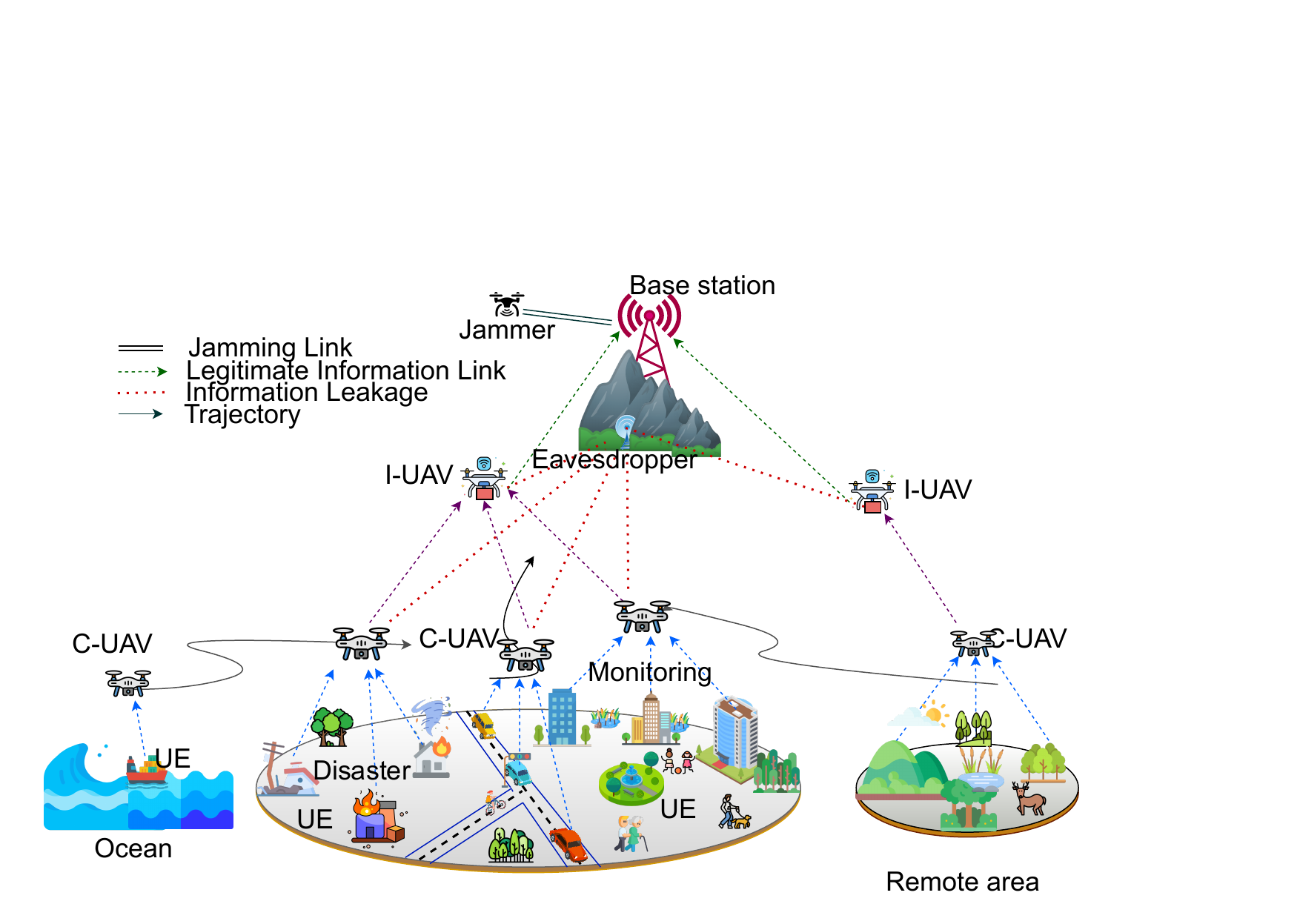}
    \caption{The system model of proposed UAV network}    
    \label{fig_1:Network}
\end{figure}

In this study, we propose an IRS-assisted AoI aware bi-layer multi-UAV system designed to enhance the reliability of wireless networks catering to multiple User Equipment (UEs) in the timely execution of computation-intensive and delay-sensitive tasks. The proposed 3D system at time stamp $t$, illustrated in Fig.~\ref{fig_1:Network}, comprises a distributed mobile user network with $M$ UEs positioned at $s_{m}^{UE}(t)$, $N$ C-UAVs deployed at $s_{n}^{C-UAV}(t)$ to provide Mobile Edge Computing (MEC) services, and a single Base Station (BS) at the origin. The system operates in the presence of a potential Jammer (J) at $s^{J}(t)$ and a single-antenna Eavesdropper (E) at $s^{E}(t)$. The exact locations of eavesdropper and jammer remain unknown. Instead, UAVs employ aerial photography target detection techniques to estimate their approximate positions, subsequently sharing this information with the BS. In our assumptions, each eavesdropper's and jammer's estimated regions, are denoted as ${s^{E}(t)}$ and ${s^{J}(t)}$, possesses a known radius $\epsilon_E$ and $\epsilon_J$ as discerned by the UAV, where $\epsilon_E \geq \|\hat{s^{E}(t)} - s^{E}(t)\|$ and $\epsilon_J \geq \|\hat{s^{J}(t)} - s^{J}(t)\|$, here $\hat{s^{E}(t)}$ and $\hat{s^{J}(t)}$ denotes the centroid of E's and J's potential region respectively. 
To overcome challenges in urban communication, we deploy $P$ IRS aided UAVs (I-UAVs). These UAVs are strategically positioned at $s_{p}^{I-UAV}(t)$ to create virtual Line-of-Sight (LoS) communication, addressing issues like data interception. Equipped with $L$ reflecting elements, the IRS, controlled by a microchip, manages communication between C-UAVs and the base station (BS). Trajectory planning for C-UAVs is crucial, considering time-varying data generation and limited energy. Coordination prevents collisions, and offloading starts when UAVs reach optimal locations. Each UE manages computation tasks periodically, with $m^{th}$ UE handling assignments represented by $W_m = [C_m, \lambda_m, D_m]$, here $D_m$ represents task data size, $C_m$ is CPU cycle count, and $\lambda_m$ is task arrival rate.. Due to limited capacity, UEs offload tasks to C-UAVs, which, constrained by size and weight, provide partial computation. C-UAVs then offload remaining tasks to the BS via I-UAVs for further processing. Our system, using Orthogonal Frequency Division Multiple Access (OFDMA), ensures simultaneous execution of tasks by multiple users, maintaining data freshness and security.

\subsection{Transmission Protocol}
The channel coefficients for various links are represented as follows: the $m^{th}$ user to $n^{th}$ C-UAV link ($h_{mn}^{UC}$), the $n^{th}$ C-UAV- eavesdropper link ($h_{n0}^{CE}$), the jammer-BS link ($h^{JB}$), the $n$th C-UAV - $p^{th}$ I-UAV link ($h_{np}^{CI} \in \mathbb{C}^{L \times 1}$), the jammer- $p$th I-UAV link ($h_{0p}^{JI} \in \mathbb{C}^{L \times 1}$), the $p$th I-UAV-BS link ($h_{p0}^{IB} \in \mathbb{C}^{1 \times L}$), and the $p$th I-UAV-E link ($h_{p0}^{IE} \in \mathbb{C}^{1 \times L}$). To model the channel properties we used Nakagami-m distribution \cite{1412063}.
Assuming accurate Channel State Information (CSI) for legitimate channels due to slow-varying nature. However, the lack of cooperation between the C-UAV and third-party nodes makes obtaining the CSI for illegitimate channels challenging, we assume partial knowledge of the CSI. So, CSI uncertainty for those channels are characterized using bounded CSI model \cite{9133130}:
\begin{align}
    h^i &= \hat{h}^i + \Delta h^i, \\ &\quad \|\Delta h_i\| \leq \xi_{h,i}, \quad i \in \{CE, JB, JI, IE\} \nonumber
\end{align}
where $\hat{h}_i$ represents the estimated CSI known at the C-UAV, and $\Delta h^i$ and is the unknown CSI error. Additionally, $\xi_{h,i}$ represents the levels of CSI uncertainty.

\begin{itemize}
    \item \textbf{User to C-UAV transmission:} The received signal at $n$th C-UAV from $m$th UE can be expressed as
\begin{align}
    y_{C-UAV}^{n} = \sum_{m=1}^{M}\frac{h_{mn}^{UC}}{\sqrt{L_{mn}^{UC}}} \cdot x_{UE}^{m}+n_{C-UAV}^{n}
\end{align}
where $x_{UE}^{m}$ is the transmitted symbol with energy $P_m$ from $m^{th}$ UE, and $n_{C-UAV}^{n} \sim N(0,\sigma_{C_n}^2)$ denotes the AWGN term with zero mean and variance $\sigma_{C_n}^2$. $L_{mn}^{UC} = A{\|s_{m}^{UE}(t) - s_{n}^{C-UAV}(t)\|}^{\alpha_{pl}}$ represents the path loss where $A$ is the constant associated with the signal frequency and transmission environment, and $\alpha_{pl}$ denotes the path loss exponent. The instantaneous received SNR at $n^{th}$ C-UAV receives signal from $m^{th}$ UE will be,
\begin{align}
\Gamma_{nm}^{C-UAV}(t) = \frac{{h_{mn}^{UC}}^2.P_m}{(\sigma_{C_n})^2+{\sum_{i=1,i\neq m}^{M}L_{in}^{UC}h_{in}^{UC}}^2.P_i}
\end{align}
In the process of offloading tasks, it is assumed that the up link bandwidth $B_u$ is distributed equally among all UEs. As a result, the instantaneous data rate between $m^{th}$ UE, and $n^{th}$ C-UAV is determined as,
\begin{align}\label{18}
      R_{mn}^{C-UAV}(t) = \frac{B_u}{M_n(t)}\log_2({1+\Gamma_{nm}^{C-UAV}})
\end{align}
here $M_n(t)$ represents the no. of active UEs served by $n^{th}$ C-UAV at time instance $t$.

\item \textbf{C-UAV to BS transmission via I-UAV:} The desired signal from $n^{th}$ C-UAV to BS with power $P_n^t$ for further computation via $p^{th}$ I-UAV is denoted as $\hat{y}_{C-UAV}^{np}$. The jamming signal $x_J$ is transmitted to the BS with power $P_J$, undetectable by the C-UAV. Each IRS element reflects a combined signal to both the BS and the eavesdropper. The $p^{th}$ IRS's reflection coefficient matrix is $\mathbf{v_p} = (v_{1p}, \ldots, v_{Lp})^T$, where $v_{ip} = e^{j\theta_{ip}}$ and $\theta_{ip} \in [0,2\pi]$ with $|v_{ip}| = 1$ for all $i$. Potential collaboration between the jammer and eavesdropper nullifies the eavesdropper's reception of the jamming signal. Multiple reflections by the IRS are negligible due to significant path loss. Consequently, the received signals at the BS and the eavesdropper by $n^{th}$ C-UAV are expressed as
\begin{align}
    y_{n}^{BS} &= \sum_{p=1}^{P}
    \frac{h^{IB}_{p0}}{\sqrt{L^{IB}_{p0}}}\cdot diag(\bold{v_p}) \cdot \frac{h_{np}^{CI}}{\sqrt{L^{CI}_{np}}} \cdot \hat{y}_{C-UAV}^{np}  \nonumber \\ & + \sum_{p=1}^{P}\frac{h^{IB}_{p0}}{\sqrt{L^{IB}_{p0}}} \cdot diag(\bold{v_p}) \cdot \frac{{h_{0p}^{JI}}^T}{\sqrt{L^{JI}_{0p}}}\cdot x_J \nonumber \\ & + \frac{h^{JB}}{\sqrt{L^{JB}}} \cdot x_J + n_{BS}\\
    y^E_n &= \sum_{p=1}^{P} \frac{h^{IE}_{p0}}{\sqrt{L_{IE}^{p0}}} \cdot diag(\bold{v_p}) \cdot \frac{h_{np}^{CI}}{\sqrt{L_{np}^{CI}}} \cdot \hat{y}_{C-UAV}^{np} + n_{E}
\end{align}
here $n_{BS} \sim N(0,\sigma_{BS}^2)$ and $n_{BS} \sim N(0,\sigma_{BS}^2)$. The instantaneous received SNR at BS and eavesdropper receiving signal from $n^{th}$ C-UAV via $p^{th}$ I-UAV will be,
\begin{align}
\Gamma_{np}^{BS}(t,v_p) &= \frac{({\Tilde{h}^{IB}_{p0} \cdot diag(v_p) \cdot \Tilde{h}_{np}^{CI}})^2.P_n^t}{(\sigma_{BS})^2+ {\Tilde{h}^{{JB}^2}} \cdot P_J+ \Phi_1} \\ 
\Gamma_{np}^{E}(t,v_p) &= \frac{({\Tilde{h}^{IE}_{p0} \cdot diag(v_p) \cdot \Tilde{h}_{np}^{CI}})^2 \cdot P_n^t}{(\sigma_{E})^2+ \Phi_2}
\end{align}

 where $\Tilde{h}$ is path loss incorporated channel coefficient,  
 
 $\Phi_1 = {\sum \sum {[(\Tilde{h}^{IB}_{p0}).diag(\bold{v_p}).\Tilde{h}_{np}^{CI}]}^2 \cdot P_n^t} + \sum {[(\Tilde{h}^{IB}_{p0}).diag(\bold{v_p}).\Tilde{h}_{0p}^{JI}]}^2 \cdot P_J $ , 
 and $\Phi_2 = {\sum \sum {[\Tilde{h}^{IE}_{p0} \cdot diag(v_p) \cdot \Tilde{h}_{np}^{CI}}^2 \cdot P_n^t]}$.
 The instantaneous data rate between $n^{th}$ C-UAV and BS having bandwidth $B_{BS}$ via $p^{th}$ I-UAV given as,
 \begin{align}\label{18}
      R_{np}^{BS}(t,v_p) & = \frac{B_{BS}}{N}\log_2({1+\Gamma_{np}^{BS}(t,v_p)})
      \\ R_{np}^{E}(t,v_p) & = \log_2({1+\Gamma_{np}^{E}(t,v_p)})
\end{align}
\end{itemize}
Secrecy rate of $n^{th}$ C-UAV data at BS will be given as,
\begin{align}
      R_{sec,n}(t,v_p) = \sum_{p=1}^{P} \max \{0,{[R_{np}^{BS}(t,v_p) - R_{np}^{E}(t,v_p)]}\}
\end{align}
\subsection{Task Offloading Protocol}
The overall task offloading process from
UEs to BS has three phases. In the first phase, UEs communicate with C-UAVs. Second phase is subdivided into two parts, i.e.
computation of tasks at C-UAVs, and simultaneously transmission from C-UAVs to BS via I-UAVs. Finally, computation at BS.

\subsubsection{Transmission from UE to C-UAV (U2C)}
The time delay, and energy consumption associated with transmission between $m^{th}$ UE and $n^{th}$ C-UAV will be expressed as,
\begin{align}
    T_{mn}^{U2C} &= \frac{D_m}{R_{mn}^{C-UAV}(t)} \\
    E_{mn}^{U2C} &= {h_{mn}^{UC}}^2 \cdot P_m \cdot T_{mn}^{U2C}
\end{align}
\subsubsection{Computation at C-UAV}
Once C-UAVs recieve all data from the UEs, each C-UAV decides the amount of task that can be computed locally. The proportion of task of $m^{th}$ UE computed at $n^{th}$ C-UAV is given as $\beta_{m0}^n \in [0,1]$, and the task executed at base station transmitted via $p^{th}$ I-UAV is given by $\beta_{mp}^n \in [0,1]$, such that
\begin{align}\label{21}
     \beta_{m0}^n+\sum_{k=1}^K\beta_{mk}^n = 1, \forall n
\end{align}
The computation delay at $n^{th}$ C-UAV while handling the task of $m^{th}$ UE is given by
\begin{align}\label{22}
     T^{C-UAV}_{mn}(t) = \frac{\beta_{m0}^nD_mC_m}{f_{mn}(t)}
\end{align}
$f_{mn}(t) = \frac{F_u}{M_n(t)}$ represents the computational resources of \textit{C-UAV n} allocated to $m^{th}$ UE, where $F_u$ is the computational resource of each C-UAV allocated equally to every UE.

Next, by taking into account the computation time and power consumption. The energy consumed by $n^{th}$ C-UAV while handling the task of $m^{th}$ UE.
\begin{align}\label{23}
     E^{C-UAV}_{mn}(t) = \kappa[f_{mn}(t)]^3{T_{mn}^{C-UAV}(t)}
\end{align}
where $\kappa$ stands for effective switched capacitance \cite{8607062}.
\subsubsection{Transmission from C-UAV to BS via I-UAV (C2I)}
Taking into account that some tasks are offloaded by a particular $n^{th}$ C-UAV for further computing to BS via $p^{th}$ I-UAV. The time delay and energy consumption for this transmission will be given as
\begin{align}
    T_{np}^{C2I}(t,v_p) & = \frac{\beta_{mp}^n \cdot D_m}{ R_{np}^{BS}(t,v_p)} \\
    E_{np}^{C2I}(t,v_p) & = T_{np}^{C2I}(t,v_p)[({h^{IB}_{p0} \cdot diag(v_p) \cdot h_{np}^{CI}})^2 \cdot P_n^t]
\end{align}

\subsubsection{Computation at BS}
After task data is offloaded by C-UAVs to BS, the BS starts processing the computation task. The computation delay at the BS  is determined in terms of task ratio $\beta_{mp}^{n}$ as
\begin{align}
     T^{BS}_{mnp}(t) = \frac{\beta_{mp}^nD_mC_m}{f_{n}(t)}
\end{align}
where $f_{n}(t) = \frac{F_{BS}}{N}$ is the computational resource allocated to $n^{th}$ C-UAV at BS. $F_{BS}$ is the resource available at BS which is divided equally among all C-UAV.

It is assumed that each C-UAV has distinct computation and communication units. As a result, computations can be executed concurrently with the transmission of tasks. So, the computational time of $n^{th}$ C-UAV will be,
\begin{align}\label{29}
\nonumber \tau^n_{t,comp} = &\sum_{m=1}^{M}[ T_{mn}^{U2C} 
\\&+ max\{T_{mn}^{C-UAV}(t) , T_{mnp}^{C2I}(t) + T_{mnp}^{BS}(t)\} ] 
\end{align}

\subsection{UAV movement}

A task is divided into $T$ timeslots of length $\tau$, denoted as $\{0, 1, \ldots, t, \ldots, T - 1\}$, where $t$ is the index of the current timeslot. Initially, all UAVs (C-UAVs and I-UAVs) are deployed at the origin. In each timeslot, an $m^{th}$ UE either generates a data packet $D_m$ or remains inactive. Data leaves the queue only when C-UAVs approach and collect it.
Each C-UAV $n$ spends $\tau_{t,move}^{n}$ moving in direction $\rho_{t}^{n} \in (0, 2\pi)$ at a fixed speed. During remaining time $\tau_{t,comp}^n$, C-UAVs collect, compute, and transmit data. Similarly, I-UAVs take $\tau_{t,move}^{p}$ to reach a location and then hover. UEs transmit complete packets each time. Each UAV has limited energy reserve $E_{\text{max}}$; the task fails if any UAV runs out of energy.
The energy consumption of $i^{th}$ UAV due to flying and hovering in timeslot $t$ will be,
\begin{align}
    E_{i}^{UAV}(t) &= \alpha_{\text{move}} \tau_{t,move}^{i} + \alpha_{\text{hover}} [\tau - \tau_{t,move}^{i}]
\end{align}
where $\alpha_{\text{move}}$, and $\alpha_{\text{hover}}$ are energy consumption coefficient while UAV is moving and hovering respectively. These coefficients are calculated using \cite{mozaffari2019tutorial}.
\begin{equation}
    \alpha = c_1\left(1 + \frac{3v_{\text{uav}}^2}{v_{\text{tip}}^2}\right) + c_2\left(\sqrt{1 + \frac{v_{\text{uav}}^4}{4v_0^4} - \frac{v_{\text{uav}}^2}{2v_0^2}}\right) + \frac{1}{2}c_3v_{\text{uav}}^3
\end{equation}
where $v_{\text{uav}} = v_{\text{move}}$ or $v_{\text{uav}} = 0$ for $\alpha_{\text{move}}$ and $\alpha_{\text{hover}}$. Constants $c_1$, $c_2$, $c_3$ depend on power, rotors, and air density. $v_{\text{tip}}$ is tip speed and $v_0$ denotes average velocity induced by the rotor.

\section{Problem Formulation and Transformation}
Initially, we present the metrics used in this paper. As previously highlighted, our specific focus revolves around data freshness and secure communication in presence of $J$ and $E$.
\begin{itemize}
\item \textbf{Threshold AoI Violation:} Let $z_m(x) \in \mathbb{R}$ be a random process in $x \in [0, X]$ represents the generation time of the oldest data at $m^{th}$ UE, where $X = T \cdot\tau$. The term $(x - z_m(x))$ signifies waiting time before data is collected. Let $G_m$ denote timeslot set where AoI threshold is violated at $m^{th}$ UE, i.e., $G_m = \{x | x \in [0, X], x - z_m(x) \geq AoI_{th}\}$. The violation ratio $\chi$ will be:
\begin{equation}
    \chi = \frac{1}{M} \sum_{m=1}^{M} \int_{0}^{X} 1_{V_p}(x) \,dx,
\end{equation}
where $1_{V_p}(\cdot)$ is the indicator function.

\item \textbf{AoI Penalty:} 
A transformation applied to AoI, where $(\gamma)$ is the soft-constrained penalty function. In this context, the expression will be given as:
\begin{equation}
    Q = \frac{1}{M} \sum_{m=1}^{M} w_m \int_{0}^{X} f(x - z_m(x)) \,dx,
\end{equation}
where weight $w_m$ is given to higher priority UEs and:
\begin{equation}
    f(x - z_m(x)) = 
    \begin{cases} 
        \Tilde{\gamma}(x - z_m(x)), & \text{if } x \in G_m \\
        x - m_p(x), & \text{otherwise}
    \end{cases} 
\end{equation}
here $\Tilde{\gamma}(x) = x + \textit{e}^x$, introduce an exponential penalty gives more priority to UEs which violates AoI threshold.  
\end{itemize}

\subsection{Problem Statement}
Our goal is to minimize the threshold AoI violation and AoI Penalty for data freshness, and maximize achievable secrecy rate for secure communication under energy and motion constraint. This will be expressed as,
\begin{align}\label{PS}
    \min_{m,p,\beta,P_n^t,s_{i,UAV}(t),v_p} Q + \chi - \sum_{n=1}^{N}R_{sec,n}(t,v_p)
\end{align}

\begin{align}
     \text{s.t.} \quad
     \text{C1:} \nonumber &\sum_{m=1}^{M}E_{mn}^{U2C}+E_{mn}^{C-UAV}+\sum_{p=1}^{P}E_{np}^{C2I}\\ \nonumber & + E_{n}^{UAV}  \leq E_{max}, \forall n \in [1,N]\\
    \text{C2:}\nonumber  &  E_{p}^{UAV} \leq E_{max}, \forall p \in [1,P] \\
    \text{C3:}\nonumber  &  \sum_{p=1}^{P}R_{np}^{E}(t,v_p) \leq R_{th}, \forall n \in [1,N]\\
     \text{C4:}\nonumber & \parallel s^{i}_{n1}(t)-s^{i}_{n2}(t) \parallel \ge D_{min} \\ \nonumber & \forall n1,n2,n1\neq n2, i \in \{C-UAV,I-UAV\}\\
     \text{C5:}\nonumber & C_{{n1}_{max}} + C_{{n2}_{max}}\le \parallel s_{n1}^{i}(t)-s_{n2}^{i}(t)\parallel \\ \nonumber &\forall n1,n2,n1\neq n2, i \in \{C-UAV,I-UAV\}
\end{align}
In our formulation, constraints C1 and C2 impose maximum energy limits on both the C-UAV and I-UAV. Subsequently, C3 sets an upper boundary on the eavesdropper's data transmission rate. Constraints C4 and C5 delineate collision and overlapping restrictions, where $D_{\text{min}}$ represents the minimum distance between two UAVs, and $C_{{n1}_{\text{max}}},C_{{n2}_{\text{max}}}$ denote the coverage radius of the two nearest UAVs.
Solving the intricate NP-hard optimization problem $($~\ref{PS}$)$ involves numerous unknowns, including the UE's location and channel conditions. UAV mobility introduces dynamism, adding complexity. Traditional optimization methods struggle with the multitude of possible solutions. To address this, the next section delves into a DRL approach, aiming to formulate a near-optimal policy with minimal environmental information. 

\subsection{Markov Decision Problem (MDP) Formulation}
To address the problem outlined above, we initially represent it as a decentralized observable MDP, denoted as \(M = \langle U, O, A, R, \Omega, \gamma \rangle\), where \(U\), \(\Omega\) and \(\gamma\) denote combine UAV set containing C-UAV and I-UAV, transition probabilities,
and the discounted factor, respectively.

\begin{itemize}
    \item \textbf{Observation Space $(O)$:} The observation space \(O\) is defined as \(\{o_t\}\). Each UAV maintains its local observation \(o_{ut}\) within a fixed sensing range, expressed as a disjoint union of \(o_{ut} (\text{env})\) and \(o_{ut} (\text{UAV})\). The former encompasses location, remaining data amount, and data generation time for all UEs within the sensing range, along with the current region of the Jammer and Eavesdropper, and jamming power (\(P_J\)). The latter includes the current position and remaining energy for all UAVs during training.
    \item \textbf{Action Set $(A)$:} The action space \(A \equiv \{a_t, v_p\}\). For each UAV, \(a_{ut} = (\rho_{t}^{u}, l_{t}^{u})\), where \(\rho_{t}^{u}\) is the angle controlling the direction of UAV movement, and \(l_{t}^{u}\) is traveling distance, bounded by maximum distance \(l_{\text{max}}\), and $v_p$ is $p^{th}$ IRS's reflection coefficient matrix for an episode.
    \item \textbf{Reward Function $(R)$:} This system incorporates two types of reward functions: intrinsic and extrinsic.
    \begin{enumerate}
    \item \textbf{Extrinsic Reward:} The environment provides the following reward for each UAV:
    \begin{align}
     R_{\text{env},u}(t) = - \{Q + \chi - \sum_{n=1}^{N}R_{\text{sec},n}(t,v_p)\}
     \end{align}
    \item \textbf{Intrinsic Reward:} This considers all penalties associated with UAV constraints:
   \begin{align}
     R_{\text{int},u}(t) = -\{\eta_1 + \eta_2 + \eta_3\}
     \end{align}
     Here, $\eta_1, \eta_2, \eta_3$ represent constant penalties applied if constraints $C_1, C_2, C_3$ are violated, respectively.
    \end{enumerate}
    In this context, the objective is to maximize the sum of overall reward $R_{u}(t) = R_{\text{int},u}(t) + R_{\text{env},u}(t)$.
\end{itemize}

\section{GTr-DRL Algorithm}

We present a novel decentralized Multi-Agent DRL (MADRL) approach for our system, integrating a Gated Transformer (GTr) for temporal modeling. Decentralized policies commonly encounter slow convergence issues. To address this challenge, we initially implement IMPALA principles \cite{espeholt2018impala} in a multi-agent context. Our proposed algorithm follows the \textit{Decentralized Training and Collaborative Execution} (DTCE) paradigm, with asynchronous actors operating on separate GPUs to enhance training efficiency.
\subsection{Transformer-Enhanced Distributed Framework}
\begin{figure}[!t]
    \centering 
    \includegraphics[width=1\linewidth]{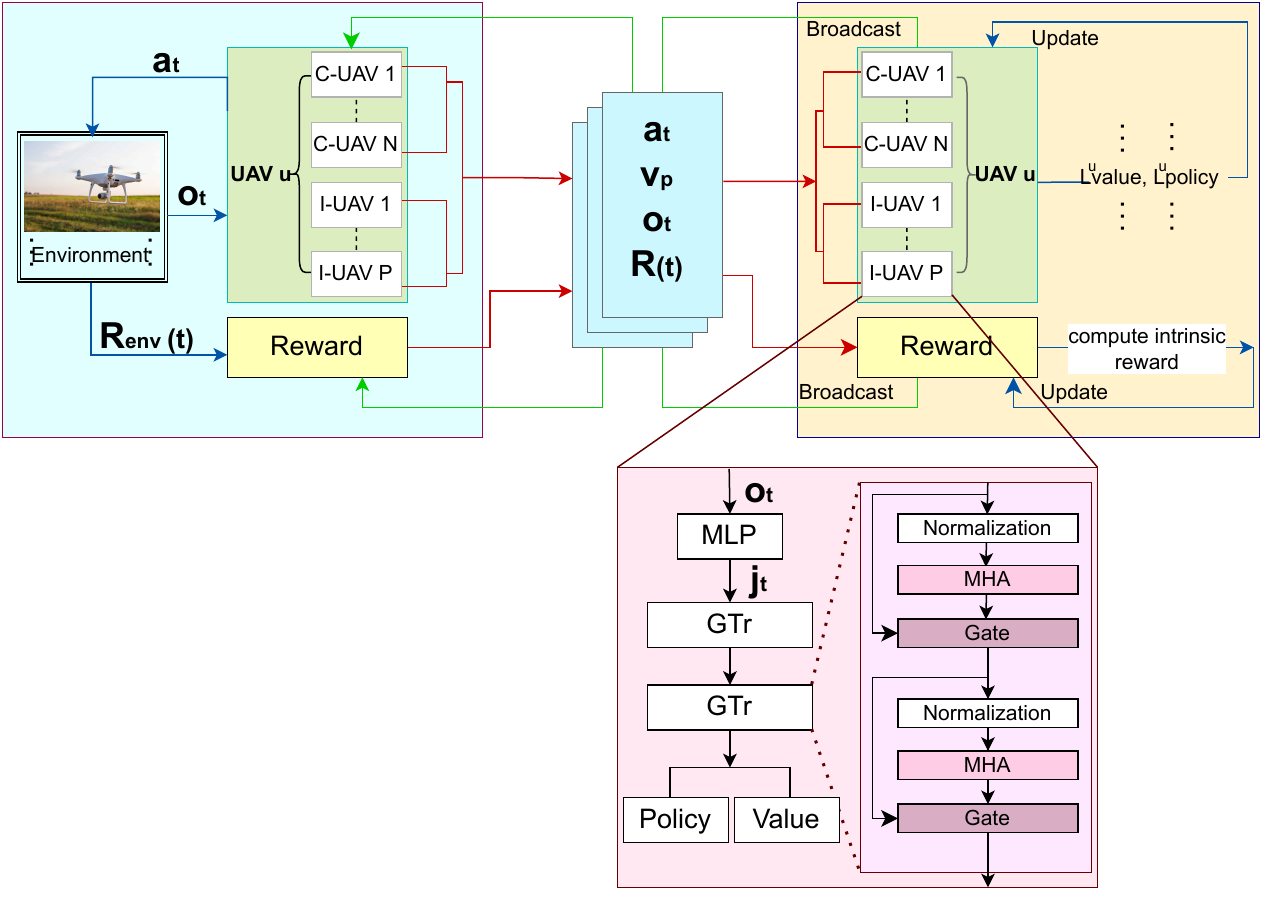}
    \caption{Proposed Solution Network}    
    \label{fig_2:Framework}
\end{figure}

The architecture incorporates GTr blocks, crucial for extracting temporal features from past trajectories. Each UAV's observation undergoes embedding via multi-layer perceptions (MLP), and temporal features are derived through multi-head attention (MHA) as shown in Fig.~\ref{fig_2:Framework}. Layer normalization is applied for model stabilization. The decentralized framework integrates policy and value networks for each UAV, addressing issues related to policy lagging and estimation variance.

V-trace target is employed to address high variance challenges in UAV optimization. Our approach utilizes \(L_2\) loss for decentralized value network optimization and policy gradients for updating the decentralized policy network. Sharing parameters between networks contributes to improved training efficiency.
The equation used for calculating the V-trace target for each UAV is given as,
\[
V_{u,t}(\mathbf{o}_u) = V_{u,\phi}(\mathbf{o}_u) + \sum_{i=t}^{T-1} \gamma^{i-t} \left(\prod_{j=t}^{i-1} c_{u,j}\right) \rho_{u,i} \cdot \text{TD}_{u,i}
\]
here $V_{u,t}(\mathbf{o}_u)$ represents the V-trace target for UAV \(u\) at time \(t\), computed based on the observation \(\mathbf{o}_u\).$V_{u,\phi}(\mathbf{o}_u)$ denotes the value estimate derived directly from the current observation for UAV \(u\). \(c_{u,j}\) corresponds to the truncated importance sampling ratio at time step \(j\). \(\rho_{u,i}\) signifies the truncated importance sampling ratio at time step \(i\). \(\text{TD}_{u,i}\) stands for the one-step temporal difference target at time step \(i\). The framework is mathematically represented with,

\begin{enumerate}
    \item \textbf{Decentralized Policy Update:}
    \begin{align}
        L_{\text{policy},u}(\theta) = E\ [&\rho_{u,t} \log \pi_{u,\theta} (a_{u,t}|o_{u,t}) \nonumber\\&(r_{u,t} + \gamma V_{u}(o_{u,t}) - V_{u,\phi}(o_{u,t}))]
    \end{align}
    \item \textbf{Decentralized Value Network Optimization:}
    \begin{align}
       L_{\text{value},u}(\phi) = E\left[(V_{u}(o_{u,t}) - V_{u,\phi}(o_{u,t}))^2\right] 
    \end{align}
\end{enumerate}

\subsection{ Optimization}
 The framework consists of independent entities called actors, operating UAVs and IRS matrices, responsible for local information gathering. A central decision-making entity, the learner, utilizes collected experiences to update deep neural network (DNN) weights, enabling decentralized and collaborative learning. 

The Actor in Algorithm~\ref{Algo1}, embodies the behavior of individual UAVs. Initialized with its unique policy network weights (\(\theta_{\text{act}}\)), episodic buffer, and other parameters, the Actor continuously interacts with the environment, gathering experiences that include UAV observations, selected actions, and environmental rewards. When the episodic buffer is full, the actor communicates with the learner by sending the experiences. If the learner broadcasts updated network weights (\(\theta\)), the actor updates its policy network accordingly.
The Learner, as depicted in Algorithm~\ref{Algo2}, functions as the central learning entity. It initializes DNN weights (\(\theta\), in a learning loop, aggregates experiences from multiple actors. The learner computes value and policy loss functions (\(L_u^{\text{value}}(\phi)\) and \(L_u^{\text{policy}}(\theta)\)) based on specific equations for each UAV. Using gradient descent methods, it minimizes the weighted sum of all losses (\(L_{\text{total}}\)). To ensure synchronization, the learner periodically broadcasts the updated network weights (\(\theta\)) to all actors. This framework adapts to diverse and dynamic scenarios through decentralized learning and collaboration.
\begin{algorithm} 
  \caption{Actor} \label{Algo1}
  \begin{algorithmic}[1]
    \Procedure{Actor}{$\theta_{\text{act}}$, episodic buffer}
      \While{learner updates}
        \State Clear episodic buffer
        \While{episodic buffer is not full}
          \State Get UAVs’ observation $o_t$ and select actions $a_t,v_p$ from policy $\pi_{\theta_{\text{act}}}$
          \State Interact with the environment and get reward $R_{u,env}(t)$
          \State Compute total rewards $R_{u}(t)$ by Equations (26) and (27) and store experiences in the buffer
        \EndWhile
        \State Send full episodic buffer to learner
        \If{received broadcast weights $\theta$}
          \State Update network $\theta_{\text{act}} \leftarrow \theta$
        \EndIf
      \EndWhile
    \EndProcedure
  \end{algorithmic}
\end{algorithm}

\begin{algorithm}
  \caption{Learner} \label{Algo2}
  \begin{algorithmic}[1]
    \Procedure{Learner}{$\theta$}
      \State Initialize network weights $(\theta)$
      \While{learner updates}
        \State Get experiences from different actors
        \State Compute $L_u^{\text{value}}(\phi), L_u^{\text{policy}}(\theta)$ for each UAV by Equations (28) and (29)
        \State Minimize weighted sum of all losses $L_{\text{total}}$ by gradient descent methods
        \If{updated times mod broadcast interval $= 0$}
          \State Send network weights $\theta$ to all actors
        \EndIf
      \EndWhile
    \EndProcedure
  \end{algorithmic}
\end{algorithm}
\section{Results and Discussion}

% \begin{figure}[!t]
%     \centering 
%     \includegraphics[width=.9\linewidth]{result.pdf}
%     \caption{C-UAV Trajectory and Task allocation}    
%     \label{fig_3: C-UAV Trajectory and Task allocation}
% \end{figure}

\begin{figure}[!t]
    \centering 
    \begin{subfigure}{0.38\linewidth}
        \centering
        \includegraphics[width=\linewidth]{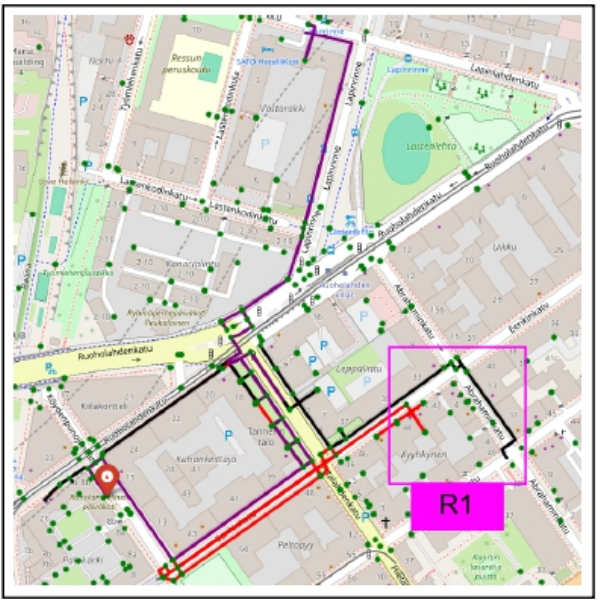}
        \caption{C-UAV Trajectory}
        \label{subfig:cuav_trajectory}
    \end{subfigure}
    \begin{subfigure}{0.48\linewidth}
        \centering
        \includegraphics[width=\linewidth]{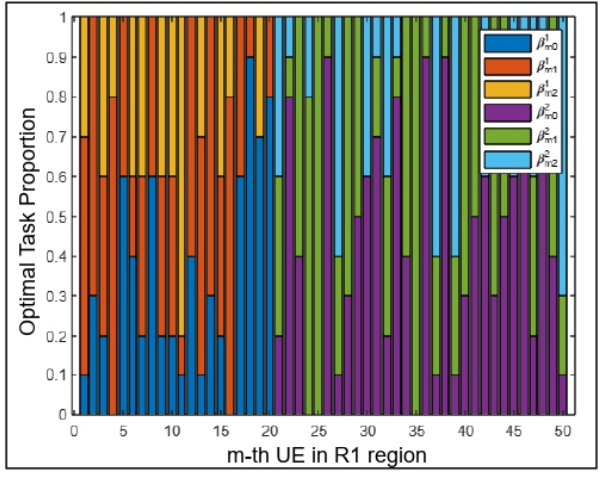}
        \caption{Task Allocation}
        \label{subfig:task_allocation}
    \end{subfigure}
    \caption{C-UAV Trajectory and Task Allocation}
    \label{fig_3: C-UAV Trajectory and Task allocation}
\end{figure}

% \begin{figure*}[!t]
%     \centering 
%     \includegraphics[width=0.68\linewidth]{r1.pdf}
%     \caption{Performance parameters vs number of C-UAVs}    
%     \label{fig_4:Performance parameters vs number of C-UAVs}
% \end{figure*}

\begin{figure*}[!t]
    \centering
    \begin{subfigure}[b]{0.22\linewidth}
        \includegraphics[width=\linewidth]{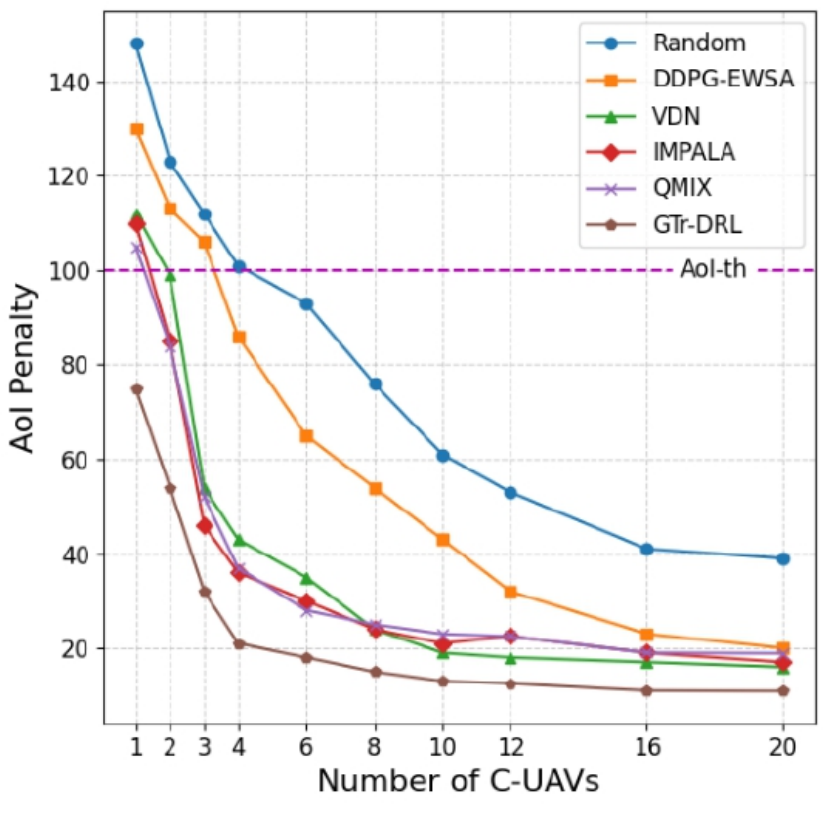}
        \caption{AoI Penalty}
        \label{fig:subfig1}
    \end{subfigure}
    \begin{subfigure}[b]{0.22\linewidth}
        \includegraphics[width=\linewidth]{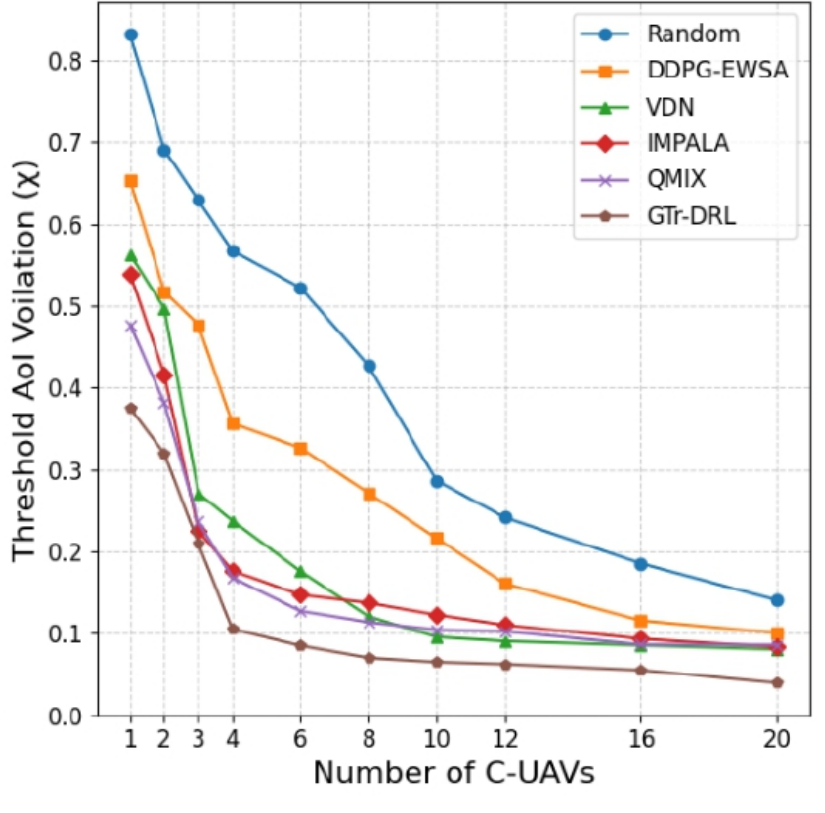}
        \caption{Threshold AoI Violation}
        \label{fig:subfig2}
    \end{subfigure}
    \begin{subfigure}[b]{0.22\linewidth}
        \includegraphics[width=\linewidth]{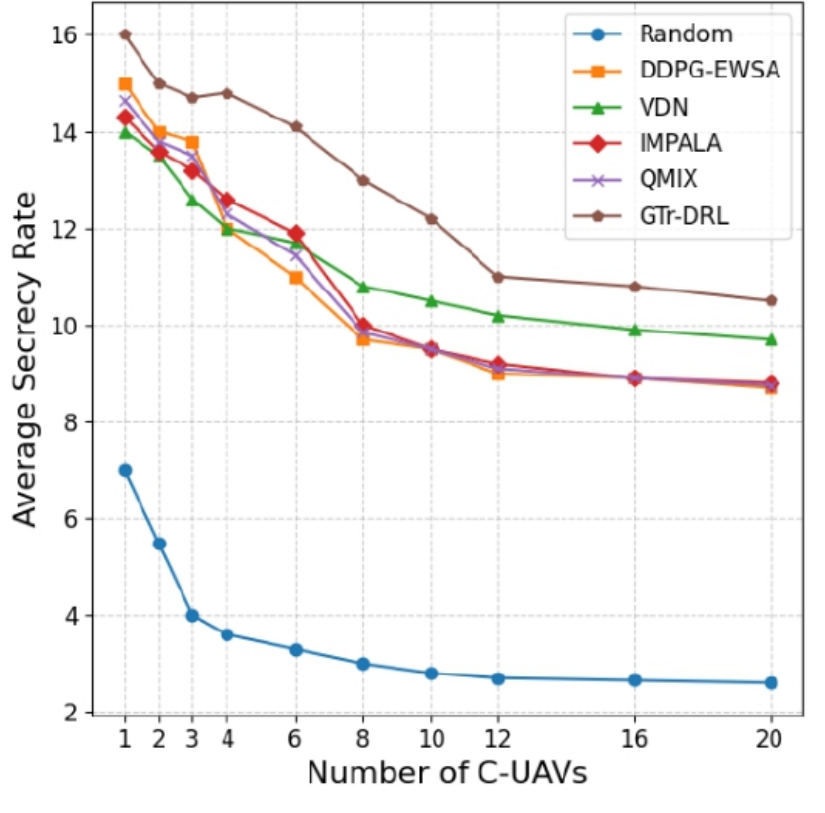}
        \caption{Average Secrecy Rate}
        \label{fig:subfig3}
    \end{subfigure}
    \caption{Performance parameters vs number of C-UAVs}
    \label{fig_4:Performance parameters vs number of C-UAVs}
\end{figure*}

\begin{figure*}[!t]
    \centering
    \begin{subfigure}[b]{0.22\linewidth}
        \includegraphics[width=\linewidth]{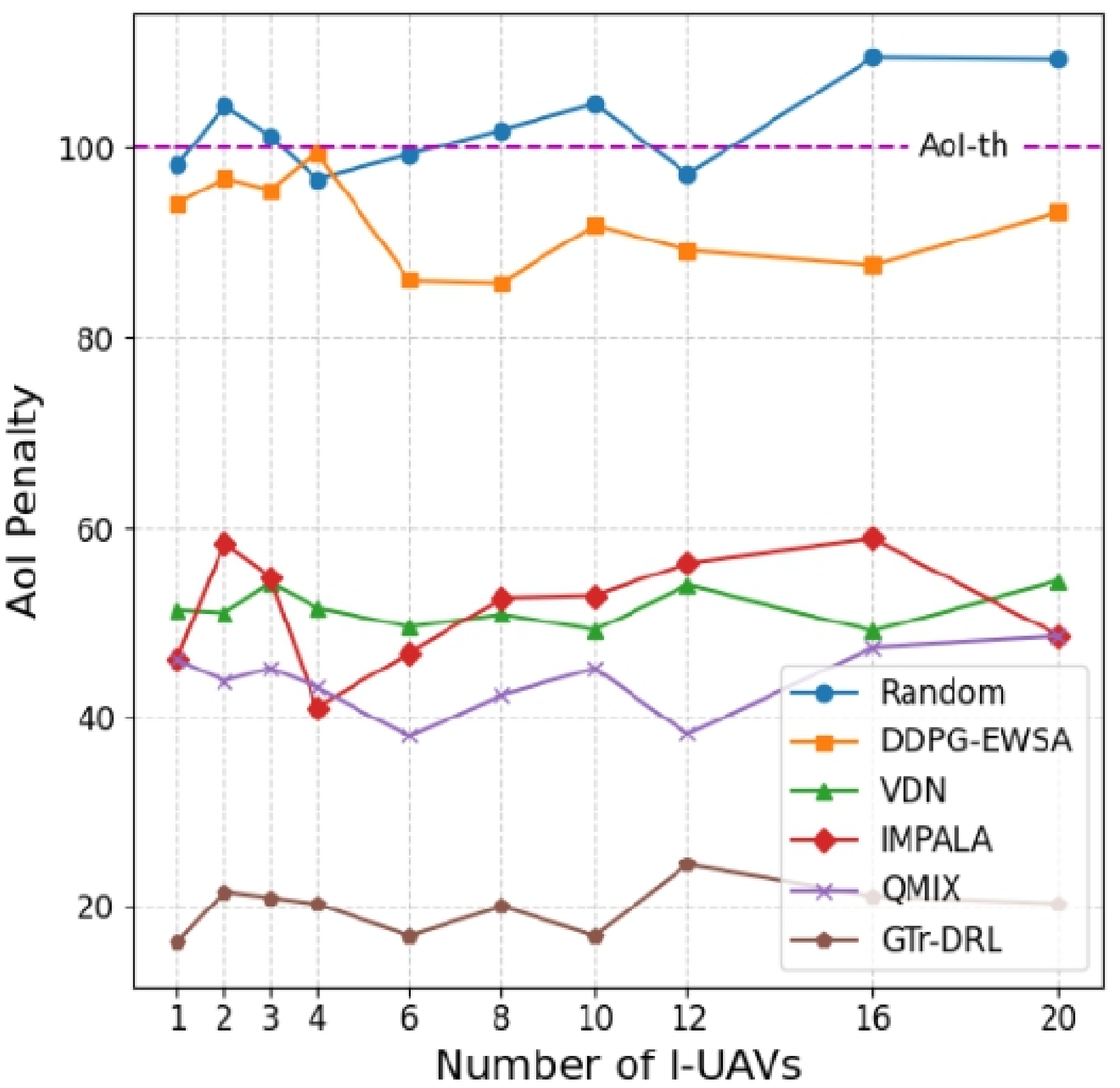}
        \caption{AoI Penalty}
        \label{fig:subfig1}
    \end{subfigure}
    \begin{subfigure}[b]{0.22\linewidth}
        \includegraphics[width=\linewidth]{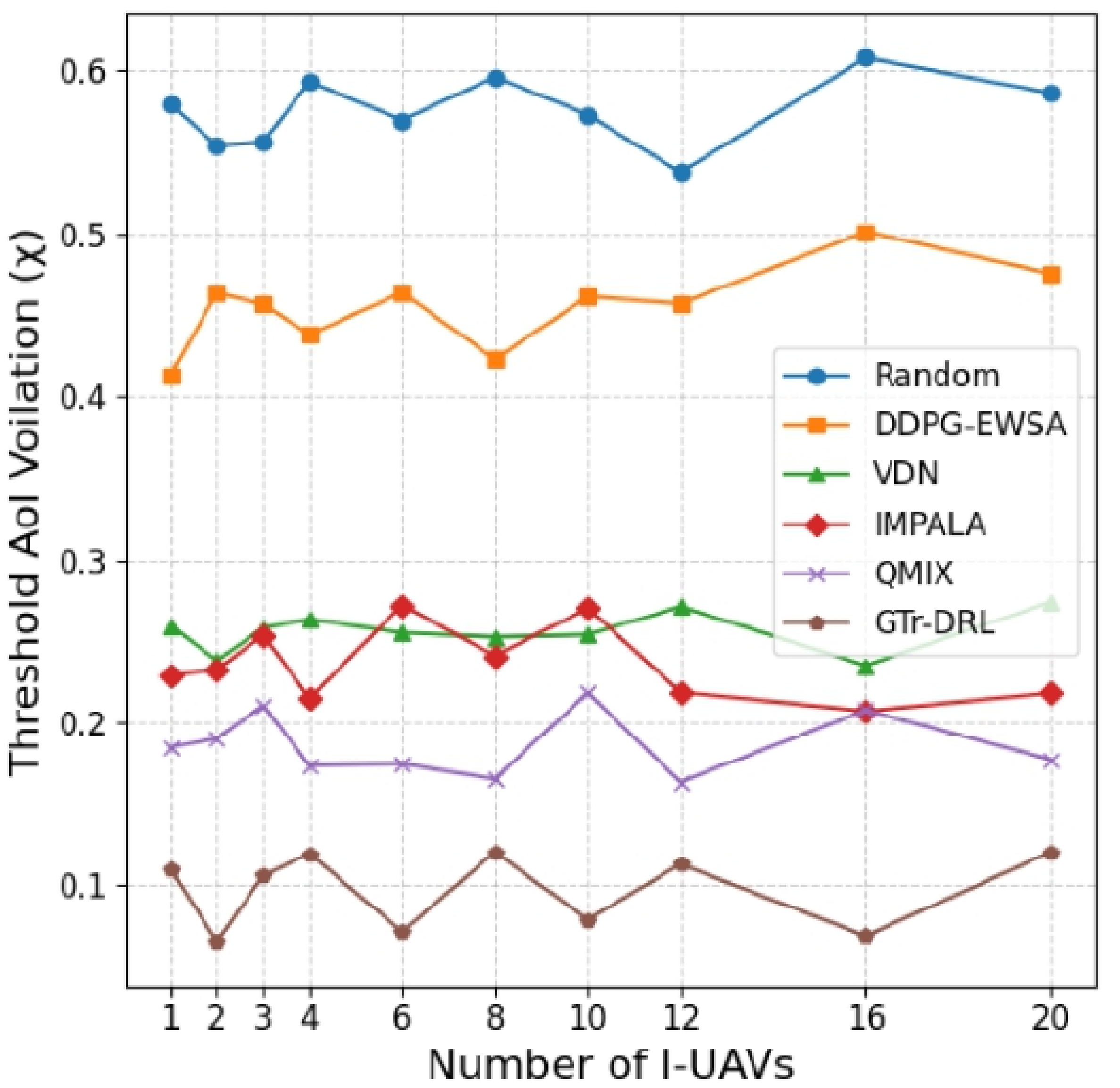}
        \caption{Threshold AoI Violation}
        \label{fig:subfig2}
    \end{subfigure}
    \begin{subfigure}[b]{0.22\linewidth}
        \includegraphics[width=\linewidth]{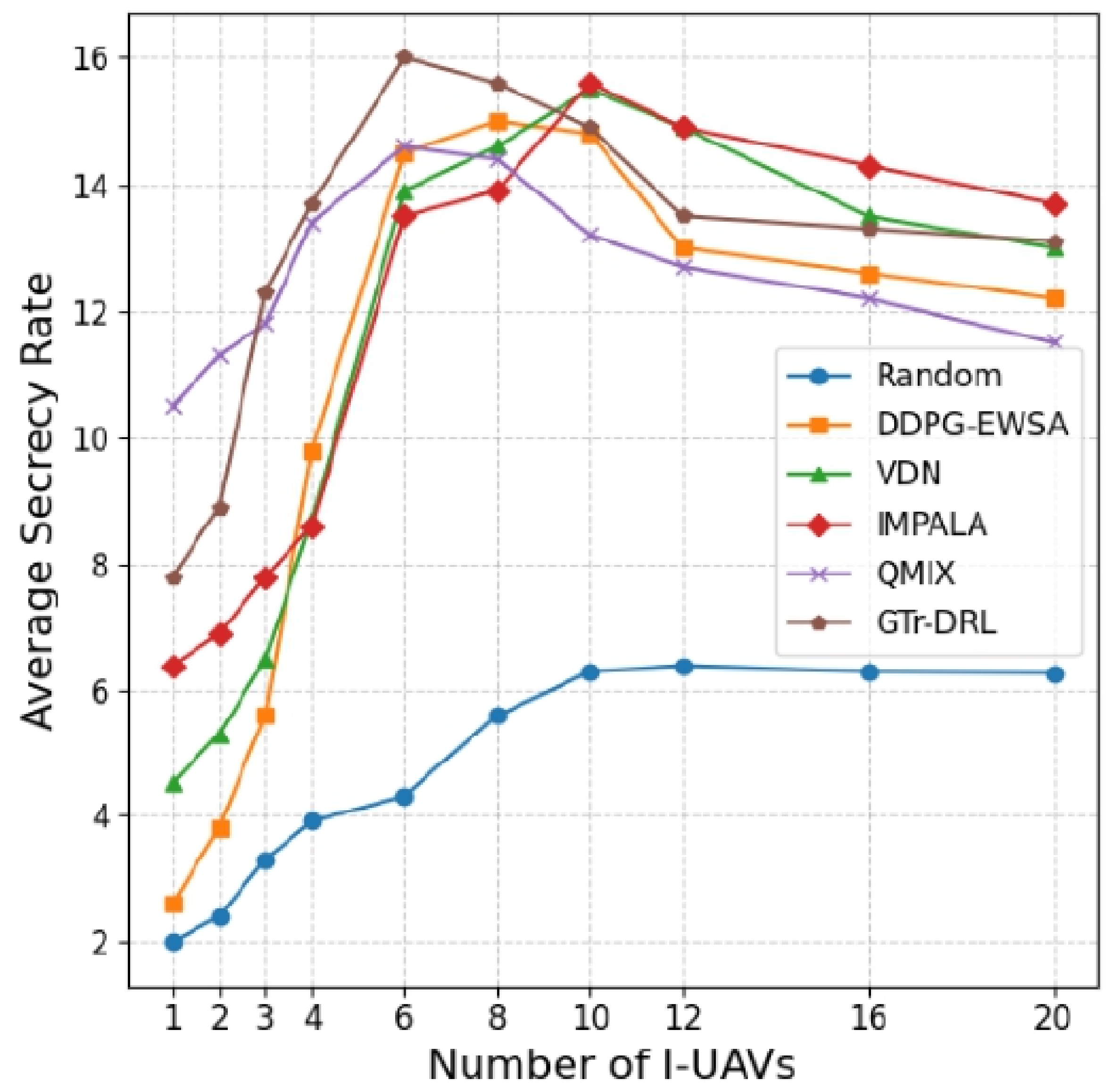}
        \caption{Average Secrecy Rate}
        \label{fig:subfig3}
    \end{subfigure}
    \caption{Performance parameters vs number of I-UAVs}
    \label{fig_5:Performance parameters vs number of I-UAVs}
\end{figure*}
% \begin{figure*}[!t]
%     \centering 
%     \includegraphics[width=0.68\linewidth]{r2.pdf}
%     \caption{Performance parameters vs number of I-UAVs}    
%     \label{fig_5:Performance parameters vs number of I-UAVs}
% \end{figure*}

In this section, we present comprehensive simulation results, including the trajectory of C-UAVs, the optimized task allocation between C-UAVs and BS via I-UAVs within our proposed learning-oriented approach. Additionally, we offer insights into key data freshness metrics, such as threshold AoI violation and AoI Penalty, alongside the average secrecy rate achieved by our learning-based scheme. These results are compared against five benchmark schemes, namely DDPG-EWSA\cite{rashid2020monotonic}, IMPALA\cite{espeholt2018impala}, VDN \cite{sunehag2017value}, QMIX\cite{rashid2020monotonic}, and random approach. 
In our simulation, the system operates within a designated area of $1500 \, \text{m} \times 1500 \, \text{m}$, with all UAVs initially positioned at the origin. UE distribution follows a normal distribution across the region, with UE positions generated at the outset and remaining constant throughout each time interval. Learning system parameters are set with a mini-batch size of $256$, a replay memory size of $2 \times 10^{5}$, $1 \times 10^{3}$ training episodes, a learning rate of $0.005$, a discount factor of $0.95$, and a soft update rate of $100$. Neural networks in the learning system comprise two hidden layers with ReLU activation functions, utilizing the Adam optimizer. City data, including the locations and configurations of tall buildings, is annotated using Google Maps and OpenStreetMap to ensure UAV collision avoidance.

Fig.~\ref{fig_3: C-UAV Trajectory and Task allocation} illustrates the trajectories of three C-UAVs collaborating with a total of six I-UAVs for security support. The red icon indicates the starting point of the C-UAVs. Additionally, the figure includes a task allocation plot for region $R1$, where 50 UEs are situated. In this region, two C-UAVs and two I-UAVs are tasked with specific assignments. For instance, the task allocation for the 50th UE showcases the collaborative deployment approach: 10\% of the task is handled by C-UAV 2, 20\% is transmitted to the BS via I-UAV 1, and the remaining 70\% is efficiently transferred by I-UAV 2.

Fig.~\ref{fig_4:Performance parameters vs number of C-UAVs} illustrates the simulation results in the presence of jammers and eavesdroppers, showcasing the impact on the AoI penalty, threshold AoI violation, and secrecy rate as the number of C-UAVs increases. In this scenario, we maintain a fixed AoI threshold (\(AoI_{th} = 100\)) and deploy 6 I-UAVs. Notably, the results reveal improved data freshness metrics. Fig.~\ref{fig_4:Performance parameters vs number of C-UAVs}(a) shows decrease in AoI Penalty which means the delay in data collection drops, and in Fig.~\ref{fig_4:Performance parameters vs number of C-UAVs}(b) we see that ratio of UEs which exceed AoI threshold significantly reduced, but average secrecy rate also goes down as shown in Fig.~\ref{fig_4:Performance parameters vs number of C-UAVs}(c), particularly in denser C-UAV network. The findings suggest a crucial trade-off between data freshness and security the system while determining the optimal number of C-UAVs. 

Fig.~\ref{fig_5:Performance parameters vs number of I-UAVs}, presents performance metrics concerning changes in the number of I-UAVs while maintaining a constant presence of 3 C-UAVs. Notably, there is no significant change in AOI-based metrics with varying numbers of I-UAVs depict in Fig.~\ref{fig_5:Performance parameters vs number of I-UAVs}(a) and (b). However, average secrecy rate shown in Fig.~\ref{fig_5:Performance parameters vs number of I-UAVs}(c), increases with increase in I-UAVs, providing enhanced support. But beyond a certain point, the average secrecy rate begins to decline. This decline is due to increase in unknown channels interacting with third-party nodes. This shows trade-off between increasing I-UAV support for security and mitigating potential risks associated with additional unknown channels.
Deploying more UAVs amplifies the optimization challenge in an exponentially  expanding solution space. Our algorithm, GTr-DRL, outshines baseline methods: DDPG-EWSA, employing DDPG with Ornstein-Uhlenbeck noise, and VDN and QMIX with $\epsilon$-greedy policies proved insufficient in this dynamic environment. IMPALA, while surpassing DDPG-EWSA, faced non-stationary problems due to centralized paradigms as neglecting specific UAVs' behavior by just relying on global joint reward. Despite enhancements, QMIX encountered hurdles from $\epsilon$-greedy exploration and lack of temporal modeling. This underscores GTr-DRL's effectiveness in multi-UAV netwrok optimisation.

\section{Conclusion}
\label{sec:conclusion}
% \textcolor{red}{Rewrite the conclusion in active voice}
In this paper, an efficient approach with IRS-assisted AoI-aware bi-layer multi-UAV system is designed to optimize task offloading and enhance overall network performance. The framework introduces exponential AoI metrics and prioritizes the maximization of secrecy rates, addressing critical challenges in UAV-assisted networks. By using multi-agent DRL, our approach efficiently allocates tasks, minimizing AoI, and ensuring robust security. The study highlights the trade-off between AoI and secrecy rate, emphasizing the need for balancing information delay and data confidentiality.

\section*{Acknowledgments}
This work is supported by the National University of Singapore under Grant No.: MoE AcRF Tier-1 FRC Grant, NUS WBS No. A-8002142-00-00

\bibliographystyle{IEEEtran}
\bibliography{ref}
  
\end{document}